\documentclass[twocolumn,showpacs,preprintnumbers,amsmath,amssymb,superscriptaddress,floatfix]{revtex4}

\usepackage{graphicx}
\usepackage{dcolumn}
\usepackage{bm}

\begin{document}


\title{Strain amplification of the 4k$_F$ chain charge instability in Sr$_{14}$Cu$_{24}$O$_{41}$}%

\author{A.~Rusydi}
\affiliation{Institut f\"{u}r Angewandte Physik,
Universit\"{a}t Hamburg, Jungiusstra$\ss$e 11, D-20355 Hamburg, Germany}

\author{P.~Abbamonte}
\affiliation{Physics Department and Frederick Seitz Materials Research
  Laboratory, University of Illinois, Urbana, IL, 61801} 
  
\author{H.~Eisaki}
\affiliation{Nanoelectronics Research Institute, AIST, 1-1-1 Central
  2, Umezono, Tsukuba, Ibaraki, 305-8568, Japan} 
  
\author{Y.~Fujimaki}
\affiliation{Department of Superconductivity, University of Tokyo, Bunkyo-ku, Tokyo 113, Japan}

\author{S.~Smadici}
\affiliation{Physics Department and Frederick Seitz Materials Research
  Laboratory, University of Illinois, Urbana, IL, 61801} 

\author{N.~Motoyama}
\affiliation{Department of Superconductivity, University of Tokyo, Bunkyo-ku, Tokyo 113, Japan}

\author{S.~Uchida}
\affiliation{Department of Superconductivity, University of Tokyo, Bunkyo-ku, Tokyo 113, Japan}

\author{Y.-J.~Kim}
\affiliation{Physics Department, University of Toronto, Toronto, Canada}

\author{M.~R\"{u}bhausen}
\affiliation{Institut f\"{u}r Angewandte Physik,
Universit\"{a}t Hamburg, Jungiusstra$\ss$e 11, D-20355 Hamburg, Germany}

\author{G.~A. Sawatzky}
\affiliation{Department of Physics and Astronomy, University of
  British Columbia, Vancouver, B.C., V6T-1Z1, Canada}

\date{\today}

\begin{abstract}

We have used resonant soft x-ray scattering (RSXS) to study the misfit strain in 
Sr$_{14}$Cu$_{24}$O$_{41}$ (SCO), a cuprate that contains
both doped spin ladders and spin chains, as well as a ``control" sample without 
holes, La$_{6}$Ca$_{8}$Cu$_{24}$O$_{41}$ (LCCO).  
The misfit strain wave in SCO is strongly temperature (T)-dependent and is 
accompanied by a substantial hole modulation.  In LCCO the strain wave is weaker, shows no 
hole modulation, and is T-independent.  
The observed strain wave vector, $L_c=0.318$, is close to the 4k$_F$ instability of the chain.  
Our results indicate that the chain charge order observed in SCO by several groups is 
a 4k$_F$ charge density wave (CDW) amplified by the misfit strain in this material.  This 
demonstrates a new mechanism for CDW formation in 
condensed matter and resolves several contraversies over the transport properties of SCO.

\end{abstract}


\maketitle

The two-leg ``spin ladder" is a simplified version of the $t-J$ model still believed by many
to contain the basic physics of high temperature superconductivity.   A doped spin 
ladder can exhibit, depending upon the parameters chosen, 
either exchange-driven superconductivity\cite{Dagotto1,Siegrist}
or an insulating ``hole crystal" (HC) ground state
in which the carriers crystallize into a static, Wigner lattice\cite{Dagotto1,White,Carr}.
The interplay between these two phases follows much of the
phenomenology of the competition between ordered stripes and superconductivity in 
two dimensions\cite{Tran1}. Detailed studies of spin ladders, both theoretical and
experimental, are therefore critical for our understanding of superconductivity in two dimensions.

This has led to a great deal of recent work on the system Sr$_{14-x}$Ca$_x$Cu$_{24}$O$_{41}$ (SCCO),
which contains both doped spin ladders and spin chains.  The ladder and chain substructures in 
this material have incommensurate periods and are mated together under internal strain.  
At $x > 10$ this system was shown to exhibit superconductivity at pressures above 
3 GPa\cite{Uehara}.  Further, it exhibits both the transport properties and elementary excitations 
characteristic of a charge density wave (CDW)\cite{Vuletic,Blumberg1}.  
Recently this system was shown, with resonant soft x-ray scattering (RSXS) studies, to contain
a commensurate Wigner crystal of holes or ``hole crystal" (HC) in the ladder 
substructure \cite{Abbamonte1,Rusydi1}.  

Prior to these studies several authors, applying neutron\cite{Matsuda} 
and hard x-ray scattering methods\cite{Fukuda,Cox}, reported a distinct modulation of the crystal structure 
at low temperature which they interpreted as a CDW in the chain substructure.  
It was later pointed out by van Smaalen\cite{Smaalen} that the observed superstructures 
index to the large unit cell of the total structure, arise naturally from the misfit strain,
and are not necessarily an indication charge order. 
It was later shown, however, that some of these reflections show anomalous temperature depedence and, regardless of their
relation to the structure, may be a sign
of charge order after all \cite{Etrillard,Zimmermann,Braden}.  The exact origin of these peaks and their 
relationship to the ladder hole crystal reported in Refs. \cite{Abbamonte1,Rusydi1} is currently a mystery.
To resolve this mystery we report here a study of the chain charge order in this system with resonant
soft x-ray scattering (RSXS). 

Single crystals of Sr$_{14}$Cu$_{24}$O$_{41}$ (SCO) and control samples of 
La$_{6}$Ca$_{8}$Cu$_{24}$O$_{41+\delta}$ (LCCO), which have no doped holes, were grown by 
traveling solvent floating zone techniques described in Ref. \cite{Motoyama}. 
The surfaces were prepared in the manner described in Ref. \cite{Abbamonte1} and the samples were
characterized {\it in situ} with x-ray absorption spectroscopy (XAS) at the O$_K$ and Cu$L_{3,2}$ edges.  
RSXS and XAS measurements were carried out on the soft x-ray undulator beamline X1B at 
the National Synchrotron Light Source using a 10-axis, vacuum-compatible diffractometer.
The energy resolution 
in the range of interest was about 200 meV. We denote periodicity with Miller indices 
of the chain, i.e. $(H,K,L_c)$ denotes a momentum transfer 
Q = ($\frac{2\pi}{a}H$,$\frac{2\pi}{b}K$,$\frac{2\pi}{c_c}L_c$) where
$\it{a}$ = 11.47 $\AA$, $\it{b}$=13.35 $\AA$, and $\it{c}$=27.3 
$\AA$$\approx$7$c_L$ $\approx$10$c_c$ \cite{Etrillard}.

\begin{figure}
\includegraphics[width=85.mm]{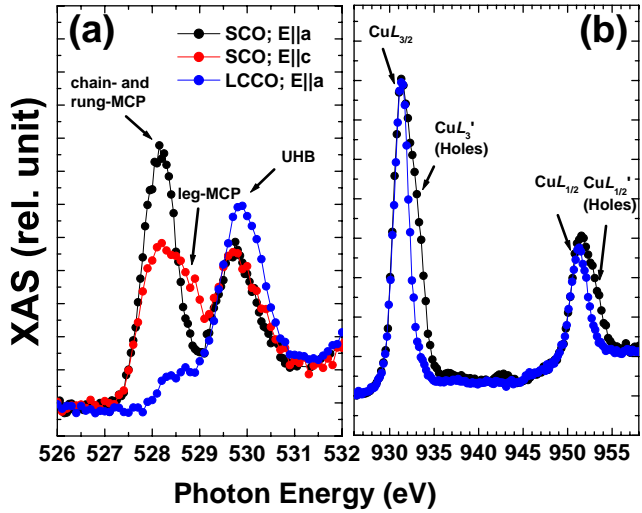}
\caption{Polarization dependent XAS of SCO and LCCO at 300 K (a) O$K$ edge
and (b) Cu$L$ edge.}
\end{figure}

Polarization-depedent XAS at the O$K$ edge 
of SCO and LCCO, taken {\it in situ} at room temperature, are shown in Fig. 1(a). 
These results are consistent with previous studies \cite{Nucker} and indicate good surface quality.  
Like the high temperature superconductors, SCO is a charge transfer insulator in the 
Zaanen-Sawaztky-Allen (ZSA) classification scheme\cite{ZSA}, therefore 
the doped holes have O2$p$ character. 
A recent reexamination of the polarization-dependent XAS \cite{Rusydi2} suggests that 
the doped holes in SCO consist of three mobile carrier peaks (MCP) - a chain MCP at 528.1 eV, 
a ladder rung MCP at 528.3 eV, and a ladder leg MCP at 528.6 eV.  The higher energy 
feature at 530 eV is assigned to transitions into the upper Hubbard band (UHB) i.e. the 
Cu3$d$ orbitals, through $p-d$ hybridization.  In contrast to SCO, in LCCO the doped hole 
features are absent which is consistent with the expectation that it is a zero hole system. 
Note that the UHB feature in LCCO is, correspondingly, stronger, indicating that spectral 
weight is transferred with doping
in this system as in other cuprates\cite{CTChen}.

Figure 1(b) shows the corresponding XAS at the Cu$L$ edge, also taken at room temperature. 
The Cu$L$ edge consists of two peaks at 931 eV and 951 eV which are transitions into the
3$d$ band from a core 2$p$ level with j=3/2 or 1/2, respectively.  
The main difference between the two samples is the presence of a 
ligand hole side band at 933 eV and 953 in SCO which is absent in the undoped LCCO. 
The spectral weight in this side band can be considered a measure of the doped
hole density.  The XAS at both the Cu$L_{3/2}$ and the Cu$L_{1/2}$ edges was found to be polarization-independent in 
the $ac$-plane for both samples. 

The first RSXS measurements were carried out on SCO at the O$K$ edge. 
In this measurement the x-ray energy was tuned to 528.4 eV at which scattering 
from the holes is enhanced by a factor of $10^3$
\cite{Abbamonte0,Abbamonte1,Abbamonte2,Rusydi1,Rusydi3}.
We first looked for a Bragg reflection at $L_c=0.20$
as was reported previously\cite{Fukuda}. 
Figure 2 shows a reciprocal space scan from $(H,K,L_c)=(0,0,0.1)$ to $(0,0,0.23)$. 
A reflection at $L_c=0.142$ is visible ($L_L = 0.2$), which is the previously reported 
hole crystal in the ladder\cite{Abbamonte1}.  
No reflection was observed near $(0,0,0.2)$.  It would also be desireable to search for 
a Bragg reflection at $L_c=0.25$ as had been reported in ref. \cite{Cox}.  However this point in 
reciprocal space cannot be reached at the O$K$ edge.  We therefore
turned to the Cu$L_{3/2}'$ edge ($\omega=933$ eV).  At this edge the chain and ladder holes 
cannot be distinguished because of its
larger radiative broadening.  However its higher energy allows larger Q values to be reached.  

Figure 2 (left) shows an $L_c$ scan from $(H,K,L_c)=(0,0,0.075)$ to $(0,0,0.4)$ at $\omega=933$ eV.
The ladder hole crystal is again observed.
No reflection was observed at either $L_c=0.25$ or 0.20.  Interestingly, however, we observed 
an incommensurate reflection at $L_c= 0.318\pm0.005$ (or $L_L=0.454$ in ladder units).  
This reflection was mapped in the $(H,L)$ plane, shown in Fig. 2 (right), giving its 
coherence lengths $\xi_c = 1100 \pm 100 \AA$ ($\sim 400$ chain units) along the
chain direction and $\xi_a= 983 \pm 100 \AA$, i.e. $\sim 87$ neighbouring chains.  
It was also observed in the zero-hole sample LCCO at $L_c= 0.300\pm0.005$.  
This reflection was 
observed previously by Hiroi {\em et. al.} \cite{Hiroi} and identified, on the assumption 
that the chain structures are more flexible than the ladders, 
as a misfit strain reflection from the chain layer.  We will accept this assignment and 
assume the modulation occurs in the chain. 

\begin{figure}
\includegraphics[width=90.mm]{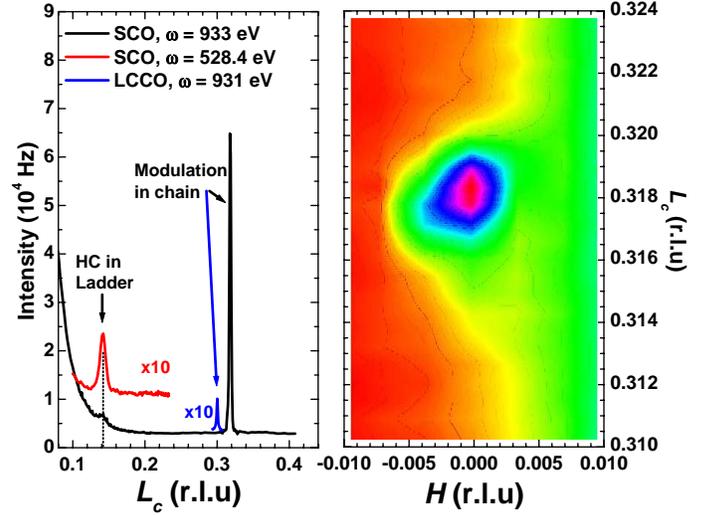}
\caption{(left) Widescans of $L$ of SCO and LCCO with different photon energies.  
 (right) HL-map of the chain modulation of SCO at T = 28 K. }
\end{figure}

\begin{figure}
\includegraphics[width=75.mm]{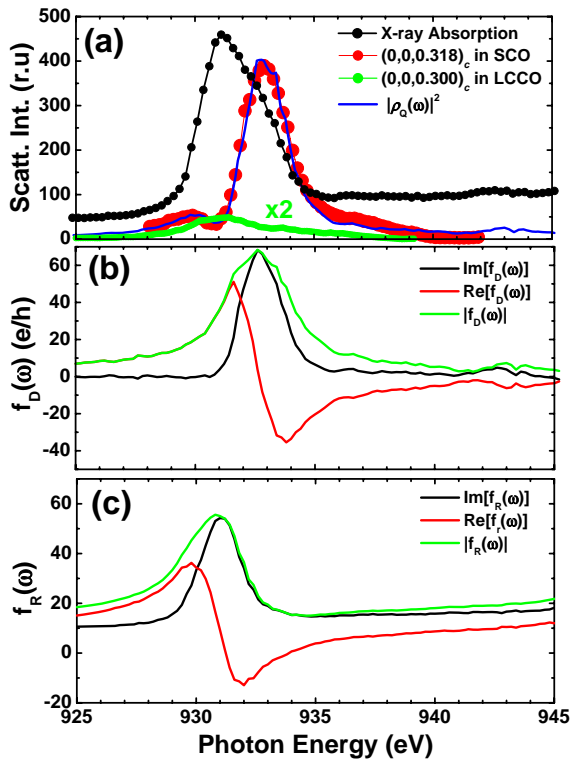}
\caption{(a) A comparison of integrated Bragg intensity with XAS of SCO and LCCO. 
Black circles, red circles and, green circles and blue line are XAS, resonance profile (RP)
of $(0,0,0.318)_c$ of SCO, RP of $(0,0,0.300)_c$ of LCCO, and scattering simulation. 
The XAS are taken at room temperature (298 K) 
with E$\parallel$a while the RP at 20 K. (b) $f_{D}(\omega)$ for hole and 
(c) $f_{R}(\omega)$ for structural (see text for detail). Spectra were placed on an 
absolute scale by fitting to tabulated values \cite{Henke} far 
from the edge.
}
\end{figure}

Our central result is the ``resonance profile" (RP) of these chain modulation peaks, 
i.e. their energy dependence through the Cu$L_{3/2}$ edge, shown in Fig. 3. These data 
were taken at $T=20K$. In SCO the RP consists of a strong resonance peak at 933 eV 
preceded by a weak resonance at 930 eV. The former coincides with the doping-dependent 
sideband and indicates the presence of a significant hole modulation \cite{Abbamonte1,Rusydi1}.  
The latter, occurring at lower energy, is likely due to the misfit strain modulation (see below).  
Evidently this chain Bragg reflection is not a simple misfit strain wave; it has a large charge 
modulation so has the character of a CDW.
By contrast, the same reflection in LCCO, also shown in Fig. 3, is much weaker and exhibits 
no hole resonance.  In this sample the chain modulation appears to be a simple strain wave.

To characterize the chain modulation we wish to quantify the relative sizes of the charge and 
strain modulations.  Because the two coexist they coherently interfere, which strongly 
influences the line shapes in Fig. 2. Recently Abbamonte \cite{Abbamonte3} 
demonstrated a method for using this interference to determine the relative size, $W$, and
phase $\phi$ of the two modulations.  In this method one divides the atomic scattering factor into
two parts, $f(\omega,v)=f_R(\omega) + v f_D(\omega)$, 
where $f_R$ is the ``raw" atomic scattering factor and $f_D$
is the scattering factor of the doped holes.  These two quantities can be determined from
doping-dependent XAS measurements, such as that in Fig. 1 (a,b), by applying the relation
	
\begin{equation}
Im[n(\omega,v)]=-\frac{r_e\lambda^2}{2\pi V_{cell}}Im[\Sigma_i f_{i}(\omega,v)],
\end{equation}

\noindent
solving a system of equations at each energy point, and applying a Kramers-Kronig transform.  
Here $Im[n]$ is the absorptive part of the refractive index (linearly related to XAS spectra), 
$r_e$ is the classical electron radius, $\lambda$ is the x-ray wavelength, and $f_i$ is the atomic scattering
factor.
The integrated intensity of the charge reflection is proportional to the quantity 
$|v_0 f_D(\omega) \exp{i\phi/2} + i Q u_0 f(\omega,<\!\!v\!\!>)|^2$, where $<\!\!v\!\!>$ is the average 
atomic valence, $v_0$ is the charge modulation amplitude,
$u_0$ is the (longitudinal) strain amplitude, $Q$ is the wave vector of the CDW, and $\phi$ is their relative phase.
A fit to the resonance profile yields values for $\phi$ and the ratio $W=v_0/u_0$.  The latter is a measure
of the degree to which the CDW is driven by interactions rather than structural effects\cite{Abbamonte3}.  

To apply this method we used the data in Fig. 1 (b) to determine $f_R(\omega)$ and 
$f_D(\omega)$ for Cu\cite{note1} in this system. Because the Cu atoms in the chains and ladders 
cannot be distinguished in XAS we simply take $<\!\!v\!\!>=0.25$ for
SCO and $<\!\!v\!\!>=0$ for LCCO and assume the resulting form factors are valid for all Cu atoms in the sample. 
The results are summarized in Figure 3.  We find that the doped-hole form 
factor $\mid$$f_D$$\mid=67$ electrons per hole on the resonance maximum.  That is, at this resonance 
a doped hole scatters as strongly as a Ho atom.

The best fit to the RP in SCO is shown in Fig 3(a).  Carrying out this fit, we find that   
the apparent two-peak structure of the RP is actually due to destructive interference between
the charge and strain waves due to the phase offset between them.  The fit yields the values
$W=1.13 \AA^{-1}$ and $\phi = -22.5^0$.  By contrast, the RP for LCCO simply
tracks its XAS spectrum.  In this case a hole amplitude is not definable and the fit yields $W\sim 0$ and $\phi$ undefined.
In ref. \cite{Abbamonte3} it was suggested that for a
conventional Peierls CDW $W \sim 1 \AA^{-1}$ whereas for a more ``exotic" CDW, such as the
static stripe in La$_{1.875}$Ba$_{0.125}$CuO$_{4}$, $W > 10 \AA^{-1}$.  
We conclude that in LCCO the chain modulation is a pure misfit strain wave.  In SCO, however,
the modulation is a cooperative amplification of the misfit strain and 
an incipient CDW with the same wave vector.  

\begin{figure}
\includegraphics[width=79.mm]{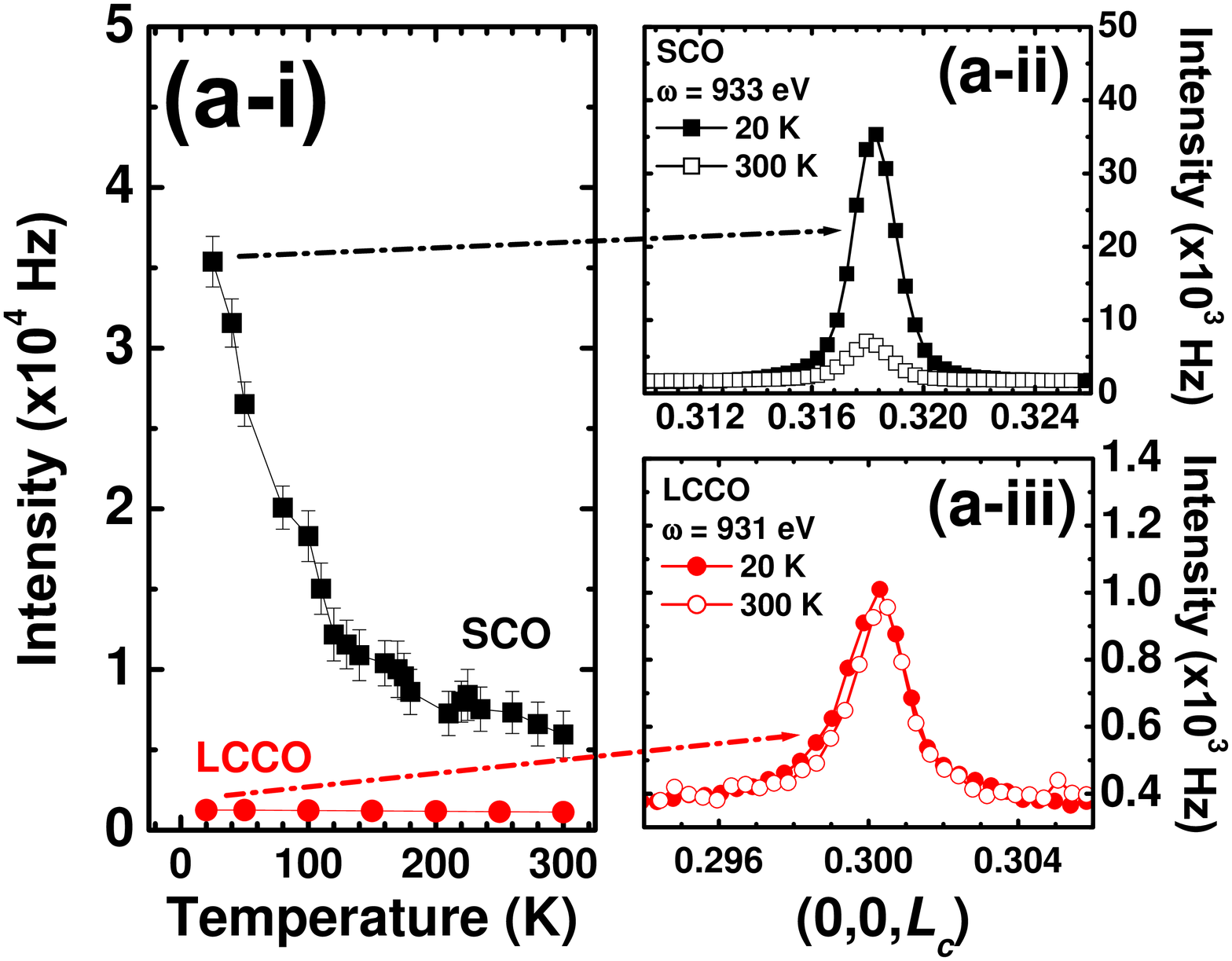}
\includegraphics[width=79.mm]{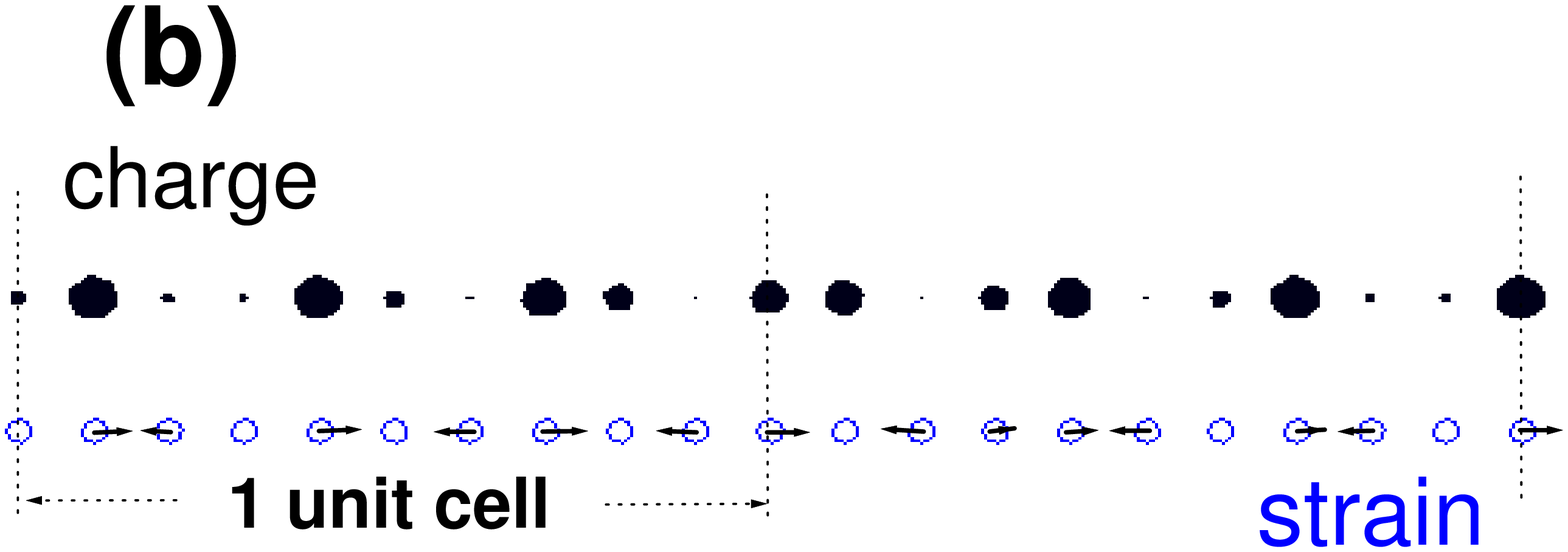}
\caption{(a-i)Temperature dependence of SCo and LCCO. (a-ii) $L$-scans for selected temperature, 
20 K and 300 K, of (a-ii) SCO and (a-iii) LCCO.
(b) Interplay between strain and charge modulation in SCO. Black and blue circles 
are the charge and strain modulation, respectively. The size of the circles represents 
the amplitude and the arrows represent the direction of the lattice distortion.
}
\end{figure}

This picture is further justified by the temperature dependence.  $L_c$ scans in both SCO and LCCO were 
carried out at different temperatures (see Fig. 4(a-i) to (a-iii)).  In SCO a pronounced 
decreased is observed upon warming, indicating that the modulation has some character of a 
CDW transition. Unlike a conventional CDW, however, the intensity never goes completely to 
zero; as the temperature rises it levels off to a constant value.  In LCCO, by contrast, 
the peak intensity is virtually temperature-independent.  This phenomenology is additional 
evidence that the chain modulation in LCCO is a simple misfit strain wave, but in SCO it 
is a cooperative effect between strain and a CDW.

In the case of SCO it was possible to determine $\phi$, therefore we can sketch the relationship 
between the charge and strain modulations (Fig. 4(b)).  We find
that the charges tend to accumulate on the ``crest" of the strain wave.  In a simple 
sinusoidal picture one might expect, by symmetry, that the charges would cluster in regions 
where the strain is either maximum or miniumum, i.e. $\phi=\pm 90^0$.
Evidently in the current case the symmetry is broken, most likely by strong anharmonicity 
of the modulation as evidenced by the large number of harmonics visible in electron 
diffraction \cite{Hiroi}.

The presence of a strain-stabilized chain CDW at higher temperature may
explain the discrepancy between results of low-frequency dielectric spectroscopy \cite{Vuletic} 
and Raman scattering \cite{Blumberg1} studies of this compound.  
Vuleti$\acute{c}$ {\em et. al.} \cite{Vuletic} found that the onset 
temperature of the CDW is about 210 K which is similar to the onset temperature 
of the HC in ladders found with RSXS \cite{Abbamonte1}.  
Blumberg {\em et. al.}\cite{Blumberg1}, however, found evidence for the presence of a 
CDW even up to the highest measured temperature, 630 K.  We propose here that the 
Raman results of Blumberg arise from fluctuations in this chain modulation, in which the 
CDW can persist to very high temperature because of the stabilizing effect of the misfit structure.

Finally, we point out an important connection between the wave vector of this reflection, 
the wave vector of the ladder hole crystal reported previuously \cite{Abbamonte1,Rusydi1}, 
and the hole count in the chains and ladders of SCO\cite{Rusydi2}.  We recently reported 
a reexamination of the hole density in this system with polarization-dependent XAS.
We concluded that, in SCO, 2.8 out of 6 holes reside in the ladders\cite{Rusydi2}, giving $\delta_L = 0.2$.  
This number is much higher than previous estimates\cite{Nucker,Osafune} and is more sensible given the 
ladder hole crystal wave vector of $L_L=0.2$ .  Here we point out that this study also implies the chain hole density is
$\delta_c = 3.2/10 = 0.32$, which is very close to the current observed wave vector of $L_c=0.318$.  
In other words the wave vector of the chain modulation we report here is very close to its $4k_F$ instablity.  
We therefore conclude that the chain modulation in SCO is an interplay between the misfit strain and the $4k_F$ instablity 
of the chain holes.

We acknowledge helpful discussions with S. van
Smaalen, M. von Zimmermann and A. Gozar. This work was supported
by the Helmholtz Association contract VH-FZ-
007, the Netherlands Organization for Fundamental Research
on Matter (FOM), and the Office of Basic Energy Sciences,
U.S. Department of Energy under Grant No. DE-FG02-06ER46285, with
use of the NSLS was supported by contract
No. DE-AC02-98CH10886.  G.A.S. was supported by the Canadian Natural Science
and Engineering Council, Foundation for Innovation, and
Institute for Advanced Research.


\begin{references}
\bibitem{note1} The scattering factors in general are tensors, however at the Cu$L_{3/2}$ edge they are isotropic in the (a,c) plane.  So we will treat them as scalars here.
\end{references}
\end{document}